\documentclass[aps,prd,showpacs,twocolumn]{revtex4}
\usepackage{graphicx}
\usepackage{hyperref}

\begin{document}

\title{Observational constraints on hyperons in neutron stars}

\author{Benjamin D. Lackey}
\author{Mohit Nayyar}
\author{Benjamin J. Owen}
\affiliation{
Center for Gravitational Wave Physics,
Institute for Gravitational Physics and Geometry,
Department of Physics,
The Pennsylvania State University,
University Park, PA 16802-6300}
\date{$$Id: paper.tex,v 1.40 2006/02/15 20:42:22 lackey Exp $$}

\preprint{IGPG-05/7-2}

\begin{abstract}

The possibility that neutron stars may contain substantial hyperon
populations has important implications for neutron-star cooling and,
through bulk viscosity, the viability of the $r$-modes of accreting neutron
stars as sources of persistent gravitational waves.
In conjunction with laboratory measurements of hypernuclei, astronomical
observations were used by Glendenning and Moszkowski [Phys.\ Rev.\ Lett.\
{\bf 67}, 2414 (1991)] to constrain the properties of hyperonic equations
of state within the framework of relativistic mean-field theory.
We revisit the problem, incorporating recent measurements of high neutron-star masses and a gravitational redshift.
We find that only the stiffest of the relativistic hyperonic equations of
state commonly used in the literature is compatible with the redshift.
However, it is possible to construct stiffer equations of state within the
same framework which produce the observed redshift while satisfying the
experimental constraints on hypernuclei, and we do this.
The stiffness parameter that most affects the redshift is not the
incompressibility but rather the hyperon coupling.
Nonrelativistic potential-based equations of state with hyperons are not
constrained by the redshift, primarily due to a smaller stellar radius.

\end{abstract}

\maketitle

\section{Introduction}

The observed masses of neutron stars have been used for years to constrain
theoretical predictions of the equation of state of degenerate matter at
high density~\cite{Glendenning:1997wn}.
If an equation of state is ``soft'' (low pressure) at high density, the
maximum mass of a stable star in general relativity is lower than for a
``stiff'' (high pressure) equation of state.
The most massive observed neutron star then sets a limit on the softness of
the equation of state.
An accumulation of neutron star-cooling observations favors (though not
decisively) the presence of exotic particles such as hyperons in the cores
of some neutron stars~\cite{Yakovlev:2004iq}, which tends to soften the
equation of state.
Hyperonic couplings in a relativistic mean-field theory of dense matter can
be constrained to a range of values based on the measured properties of
$\Lambda$ hypernuclei and the maximum neutron-star
mass~\cite{Glendenning:1991es}.
This information is in turn useful for predictions of the bulk viscosity of
hyperonic matter~\cite{Lindblom:2001hd, Haensel:2001em}, which has
important implications for the viability of $r$-modes in accreting neutron
stars as persistent sources of gravitational
radiation~\cite{Wagoner:2002vr}.
Spurred by these implications and by new observations, we revisit the
constraints on hyperonic equations of state.

First we consider the maximum mass.
The most precise observations of neutron-star masses come from radio
pulsars in binaries, which are all measured with 95\% confidence to be less
than $1.5~M_\odot$~\cite{Thorsett:1998uc}.
Accreting neutron stars are naturally expected to be more massive, and
x-ray measurements have long suggested that this is so.
The best case for decades has been Vela~X-1 (4U~0900-40), with a most
likely mass of about $1.8~M_\odot$ but with $1.5~M_\odot$ included in the
95\% confidence interval~\cite{Joss:1984}.
However, this measurement is now known to be contaminated by oscillations
of the high-mass main sequence companion~\cite{vanKerkwijk:1995ie}; and
while more recent measurements~\cite{Barziv:2001ad, Quaintrell:2003pn} can
claim smaller statistical confidence intervals, they are still subject to
large and poorly quantified systematic errors.
Thus $1.5~M_\odot$ has remained the constraint for many years.

However, this is changing.
Recent radio observations of PSR~J0751+1807 by Nice {\it et
al.}~\cite{Nice:2003tj, Nice:2004fn, Nice:2005fi} yield a neutron-star mass
greater than $1.6~M_\odot$ at the 95\% confidence level.
PSR~J0751+1807 orbits a white dwarf which, unlike the main sequence star in
Vela~X-1, shows no evidence of oscillations.
The orbital period is 6~hours, short enough that its decay due to
gravitational radiation is observable.
This provides a post-Keplerian parameter to disambiguate the two masses.
Marginal detection of the Shapiro delay implies intermediate orbital
inclination angles, and disambiguates the inclination angle (somewhat) from
the mass of the neutron stars (see the figure in Ref.~\cite{Nice:2004fn}).
Also, Ransom {\it et al.}~\cite{Ransom:2005ae} find through measurements of
the periastron advance of the highly eccentric orbit that Ter~5~I has
$1.68~M_\odot$ or higher, formally at the 95\% confidence level.
Overall this bound is tighter than that for PSR~J0751+1807, but the
companion is probably a white dwarf and there may be some contamination of
the relativistic periastron advance by its rotationally induced quadrupole.
Thus the mass constraint on equations of state is now at least
$1.6~M_\odot$ and may be $1.7~M_\odot$.

Another constraint is the measurement of a gravitational redshift by
Cottam, Paerels, and Mendez~\cite{Cottam:2002cu}.
The low-mass x-ray binary EXO0748-676 displayed several absorption lines
(inferred from multiple x-ray burst spectra) consistent with a redshift
$z=0.35$.
Estimates of numerous possible sources of error in the redshift amount to a
total of no more than 5\%~\cite{Bhattacharyya:2004nb}, implying that
equations of state should be ruled out if their maximum redshift is below
about 0.33.

Qualitatively, it has been stated~\cite{Cottam:2002cu} that softer
equations of state are disfavored by the redshift; here we make that
quantitative.
Like Glendenning and Moszkowski~\cite{Glendenning:1991es}, we constrain the
parameters of relativistic mean field theory between astronomical
observation and hypernuclear experiment.
Since there is a fairly large parameter space involved, most papers using
the results of relativistic mean-field theory use two canonical parameter
sets corresponding to a soft equation of state and a stiff equation of
state.
We find that the so-called stiff equation of state is actually the softest
allowed (marginally) by the redshift observation if hyperons are present.
However, we can and do construct stiffer equations of state that are
compatible with hypernuclear measurements and consistent with the redshift
observation.
The most important stiffness parameter as far as the redshift is concerned is
the hyperon coupling (and thus hyperon population) rather than the
incompressibility.
We also note that nonrelativistic potential-based equations of state are
not greatly constrained by the redshift observation (they can all reproduce
it due to their smaller stellar radii).
The new neutron-star masses do not constrain the relativistic mean-field
equations of state much compared to the redshift, but potential-based
models are constrained more effectively by the masses than by the redshift.
In the Appendix we provide tabulations of several relativistic mean-field
theory equations of state with hyperons that satisfy the new observational
constraints for a variety of nuclear-matter parameters.

\section{Equations of state}

We consider two types of high-density equation of state in this paper.
The first is the main focus of the paper, and the second is used for
comparison to demonstrate model dependence.

The first is based on relativistic mean-field theory and is discussed in
detail in Ref.~\cite{Glendenning:1997wn}.
(We note that other relativistic models, such as the relativistic
Br\"uckner-Hartree-Fock of Ref.~\cite{Sumiyoshi:1995cw}, produce equations of
state with qualitatively similar behavior.)
Here the low-energy strong nuclear interaction is modeled as the tree-level
exchange of mesons between baryons (neutrons, protons, and possibly
hyperons).
The starting point is the construction of a relativistic Lagrangian, which
is a sum of free-particle Dirac terms for the baryons and leptons, plus
free-particle terms for the mesons (scalar $\sigma$, vector $\omega$, and
isovector $\rho$), plus interaction terms including tree-level meson-baryon
interactions and perturbative self-interactions for the $\sigma$ meson.
This makes the theory a phenomenological low-energy effective field theory,
although it has the advantage of being many-particle and relativistic by
construction so that the sound speed never exceeds the speed of light.
It also has a small number of parameters which can be fit simply to
experiment; although this is a mixed blessing since the many numbers known
from nuclear experiments must be distilled to a few.
Mean-field theory also has the disadvantage that it neglects correlations
by construction.

Under the assumption that the bulk matter is (on a macroscopic scale)
static and homogeneous, the fields are replaced by their mean values, time
and spatial derivatives vanish, and the Euler-Lagrange equations take a
form that is relatively simple to solve but is still somewhat lengthy and
thus we do not reproduce it here.
It is enough to state that the Euler-Lagrange equations in this
approximation reduce to a set of coupled algebraic equations for the lepton
and baryon Fermi momenta and the meson fields.
These are combined with equations for generalized $\beta$-equilibrium,
electric charge conservation, and conservation of baryon number to obtain
the Fermi momenta and meson fields as functions of, for example, the total
baryon number density.
These are then used to construct the pressure and energy density, i.e.\ the
equation of state.

The Euler-Lagrange equations feature five free parameters, which under
certain assumptions are fit algebraically to numbers distilled from
laboratory measurements of many finite nuclei: the saturation density,
binding energy per nucleon and isospin asymmetry coefficient at saturation
density, and the overall incompressibility $K$ and effective mass $m^*$ of
nucleons in the nuclear medium.
The latter two are difficult to estimate from available data and are
subject to systematic uncertainties, and thus papers using this
relativistic set of equations of state typically treat a range of values
for $K$ (240--300~MeV) and $m^*$ (0.70--0.80 times the nucleon mass $m$).
(The compressibilities are typically lower for nonrelativistic models.)

At roughly twice nuclear density in these equations of state, the neutron
Fermi momentum is high enough to make hyperon production favorable in spite
of the roughly 200~MeV/$c^2$ mass difference.
The hyperons of most interest are the $\Lambda$ and $\Sigma^-$ hyperons,
which have the lowest masses and therefore are created at the lowest
densities and occupy the largest fraction of the volume of a star.
However, the other hyperons $\Sigma^0$, $\Sigma^+$, $\Xi^-$, and $\Xi^0$
also appear in small numbers at the very highest densities.
Hyperons introduce more free parameters.
The hyperon-meson couplings are assumed to be the same for all hyperons but
are weaker than the nucleon-meson couplings by the ratios $x_\sigma$,
$x_\omega$, and $x_\rho$ for the three mesons.
The former two ratios are obtained algebraically from the measured binding
of $\Lambda$ hyperons in nuclear matter (double-$\Lambda$ hypernuclei) and
(more roughly) from hypernuclear energy levels, resulting in
$x_\sigma\le0.72$ (usually taken to be $x_\sigma=0.6$) and $x_\omega$ being
determined as a function of $x_\sigma$ and $m^*$.
[Note that the first Table in Ref.~\cite{Glendenning:1991es} contains a
typo which is repeated in Ref.~\cite{Glendenning:1997wn}: The value of
$x_\omega$ reading 0.568 should read 0.658, as can be seen by solving
Eq.~(5.59) of the latter reference.]
The remaining ratio $x_\rho$ is unconstrained by hypernuclear data, since
the $\Lambda$ is isospin neutral and the relevant $\Sigma^-$ hypernuclear
measurements are highly uncertain.
As is standard practice, we set $x_\rho = x_\sigma$, although the final
equation of state is rather insensitive to the precise
value~\cite{Glendenning:1997wn}.

Numerically, we construct these equations of state using the methods of
Ref.~\cite{Glendenning:1997wn} as functions of the most uncertain
parameters $m^*$, $K$, and $x_\sigma$.
(The low density equation of state is the standard BPS
model~\cite{Baym:1971pw}, but this has little effect on the mass-radius
curve which is the subject of our work.)
For all cases we also use the values 0.153~fm$^{-3}$ for the saturation
density, -16.3~MeV for the binding energy per baryon at that density, and
32.5~MeV for the isospin symmetry energy coefficient at that density, all
as in Ref.~\cite{Glendenning:1997wn}.
We use seven fiducial equations of state of this type:
Three sets of values from Ref.~\cite{Glendenning:1997wn} are already in
common use in the literature:
$K=240$~MeV and $m^*/m=0.78$, the softest choice which we denote H1;
$K=300$~MeV and $m^*/m=0.78$, an intermediate choice denoted H2; and
$K=300$~MeV and $m^*/m=0.70$, the stiffest of these equations of state
denoted H3.
All use $x_\sigma=0.6$.
We construct the stiffest such equation of state compatible with
experimental data (H4) using $K=300$~MeV, $m^*/m=0.70$, and
$x_\sigma=0.72$.
We construct three others (H5--H7) for extreme values of $K$ and $m^*$ with
$x_\sigma$ just satisfying the astronomical constraints (see below and
Fig.~\ref{f:surface}).
These parameter values are summarized in Table~\ref{eospar}.
We also construct relativistic mean field equations of state without
hyperons by artificially setting the hyperon masses to arbitrarily high
values.
These are denoted G1--G7 correspondingly, but note that G3 is identical to
G4.

\begin{table}[b]
\caption{\label{eospar}
Parameters for seven fiducial hyperonic equations of state in relativistic
mean field theory.
Corresponding equations of state without hyperons are denoted G1--G7, but
G3 and G4 are identical.
}
\begin{ruledtabular}
\begin{tabular}{lccc}
Name & $K$ (MeV) & $m^*/m$ & $x_\sigma$
\\
\hline
H1 & 240 & 0.78 & 0.60
\\
H2 & 300 & 0.78 & 0.60
\\
H3 & 300 & 0.70 & 0.60
\\
H4 & 300 & 0.70 & 0.72
\\
H5 & 300 & 0.80 & 0.66
\\
H6 & 240 & 0.70 & 0.67
\\
H7 & 240 & 0.80 & 0.69
\end{tabular}
\end{ruledtabular}
\end{table}

The second type of equation of state is based on detailed modeling of the
potentials observed in laboratory nuclei, such as done in
Ref.~\cite{Akmal:1998cf}.
That paper, denoted APR, gives the canonical Schr\"odinger
(non-relativistic) model including detailed potentials, two- and three-body
interactions, and with some relativistic effects in the form of
perturbations.
It has the advantage of using more of the known experimental numbers than
relativistic mean-field theory, including correlations and scattering data.
However, the fitting to experimental numbers is more involved, while there
are in the end only a few numbers that characterize bulk matter and neutron-star structure.
These equations of state are few-body by construction and are fundamentally
nonrelativistic, resulting in causality violation at high densities (the
sound speed exceeds the speed of light).
Since we expect the recent observations to rule out softer equations of
state, we also consider a very soft version of this type denoted BPAL12 in
Ref.~\cite{Prakash:1996xs}.
The incompressibility of BPAL12 is 120~MeV, which was known at the time to be
much too low.
It was created explicitly to produce ``artificially'' the softest equation
of state compatible with then-known neutron star masses ($1.45~M_\odot$).
Since neither of these equations of state includes hyperons, we also
consider the results of Balberg, Lichtenstadt, and
Cook~\cite{Balberg:1998ug}, who include hyperons in a similar model, and
denote their equations of state as BLC1 (the softer) and BLC2 (the
stiffer).

\section{Maximum mass}

General relativity predicts a maximum mass for a star stable to radial
perturbations for a given equation of state.
This is seen by solving the well-known Oppenheimer-Volkoff (OV) equations,
which map a curve $p(\rho)$ (pressure as a function of energy density) onto
a curve $M(R)$ (mass of the star as a function of radius).
The gravitational mass $M$ generically has a maximum, which rules out
equations of state that are too soft to produce the observed masses.
The tightest observational constraint at 95\% confidence is now
$1.68~M_\odot$ for Ter~5~I~\cite{Ransom:2005ae}, though the corresponding
$1.6~M_\odot$ for PSR~J0751+1807~\cite{Nice:2005fi} may be cleaner.
While PSR~J0751+1807 has a 287~Hz rotation frequency~\cite{Lundgren:1995},
this is well below the mass-shedding limit for all equations of state and
can be shown to increase the OV maximum mass (which assumes no rotation) by
no more than about 2--3~percent~\cite{Cook:1993qr}.
Ter~5~I rotates at 104~Hz~\cite{Ransom:2005ae}, and thus its maximum mass
is increased by less than 1\% over the OV value.

We use a version of the OV equations due to Lindblom~\cite{Lindblom:1992}:
\begin{eqnarray}
\frac{dm}{dh} &=& -\frac{ 4\pi\rho(h)r(h)^3 [r(h)-2m(h)]} {m(h) + 4\pi
r(h)^3p(h)},
\\
\frac{dr}{dh} &=& -\frac{r(h) [r(h)-2m(h)]} {m(h) + 4\pi r(h)^3p(h)},
\end{eqnarray}
using as independent variable the specific enthalpy
\begin{equation}
h(p) = \int_0^p dp' / [p' + \rho(p')].
\end{equation}
(Here $G=c=1$.)
Unlike the standard OV equations for $m(r)$ (the mass contained within a
sphere of radius $r$) and $p(r)$, Lindblom's form does not suffer numerical
difficulties near the surface of the star, which is simply and robustly
defined by $h=0$.
We start by picking a central enthalpy and evaluating an analytical
expansion of the equations at a point very close to the center of the star
(where the equations are singular).
We then integrate down to $h=0$ and read off the total mass and radius of
the star as $M=m(0)$ and $R=r(0)$.

\begin{figure}[b]
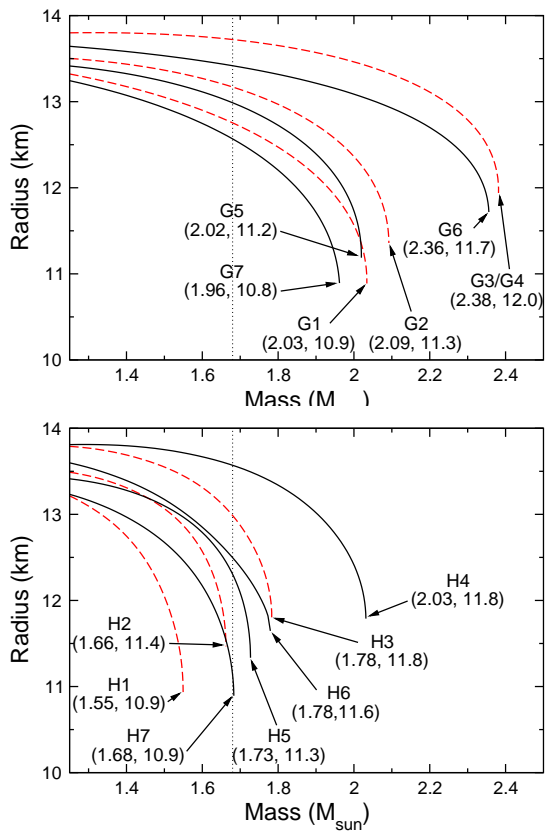

\includegraphics[width=2.8in]{RvsMnoH.eps}
\includegraphics[width=2.8in]{RvsMwithH.eps}
\caption{Oppenheimer-Volkoff mass-radius curves for the fiducial
relativistic mean field equations of state with hyperons (bottom) and
without hyperons (top).
The vertical line is the observational 95\% confidence limit $1.68~M_\odot$
from Ter~5~I.
H1 is ruled out by the observation.
H2 is marginally consistent, allowing for a small increase in maximum mass
due to rotation and the imprecision of the constraint.
}
\label{f:RvsMG}
\end{figure}

In Fig.~\ref{f:RvsMG} we plot $R(M)$ for our fiducial relativistic mean-field equations of state, with hyperons (bottom) and without hyperons
(top).
All equations of state without hyperons are consistent with $1.68~M_\odot$.
The softest one with hyperons (H1) has a maximum mass of $1.55~M_\odot$ and
is ruled out by the new pulsar observations~\cite{Ransom:2005ae,
Nice:2005fi}.
H2 is nominally inconsistent with the 95\% confidence limit of
$1.68~M_\odot$ for Ter~5~I.
However, in practice H2 cannot be ruled out by this observation and must be
considered marginally consistent because rotation (not included in the OV
model) can account for most of the $0.02~M_\odot$ difference and changing
the confidence level very slightly from 95\% would bring it within the
limit.
At the 68\% confidence level for PSR~J0751+1807~\cite{Nice:2005fi}
($1.8~M_\odot$), only the stiffest equation of state with hyperons (H3) is
marginally allowed, while again all equations of state without hyperons are
allowed.

Of the potential-based equations of state (plotted in~\cite{Akmal:1998cf,
Balberg:1998ug}), BPAL12 is firmly ruled out with a maximum mass of
$1.45~M_\odot$ (at a radius of 9.0~km).
This is not too surprising, since BPAL12 was deliberately constructed with
an artificially low incompressibility $K=120$~MeV as an extreme example.
The extremely soft example for this type of equation of state should now be 
BPAL21, which has a maximum mass of $1.67~M_\odot$ for a nonrotating star
(the 95\% confidence limit from Ter~5~I) at a radius of 9.2~km.
The hyperonic BLC1 equation of state has a maximum mass of $1.55~M_\odot$,
which is also firmly ruled out.
The stiffer BLC2 has a maximum mass of $1.75~M_\odot$, which is compatible
with the 95\% confidence limit of Ter~5~I.
APR stars have a maximum mass of $2.2~M_\odot$, compatible with all
constraints.

\section{Gravitational redshift}

General relativity also predicts a redshift for photons leaving the surface
of a star with a strong gravitational field.
For a nonrotating star the redshift $z$ obeys the relation
\begin{equation}
1+z = \left(1-\frac{2GM}{c^2R}\right)^{-1/2}.
\end{equation}
Since $R$ decreases with $M$ as $M$ approaches its maximum for a stable
star, $z$ generically has a maximum for the maximum-mass star and an
observation can rule out equations of state which cannot produce a strong
enough redshift.
Cottam, Paerels, and Mendez~\cite{Cottam:2002cu} have such an observation,
a gravitational redshift $z=0.35$ obtained by identifying several
absorption lines in spectra constructed from multiple bursts from
the low-mass x-ray binary EXO0748-676.
Due to the number of consistent lines the result is robust, although there
may be errors at the few percent level~\cite{Bhattacharyya:2004nb}.
Pulsations from a more recent x-ray burst have inferred a rotation
frequency of 45~Hz for the neutron star~\cite{Villarreal:2004nj}.
At this frequency rotational corrections to the redshift should be a
fraction of a percent~\cite{Bhattacharyya:2004nb} and the nonrotating
approximation suffices.

\begin{figure}[t]
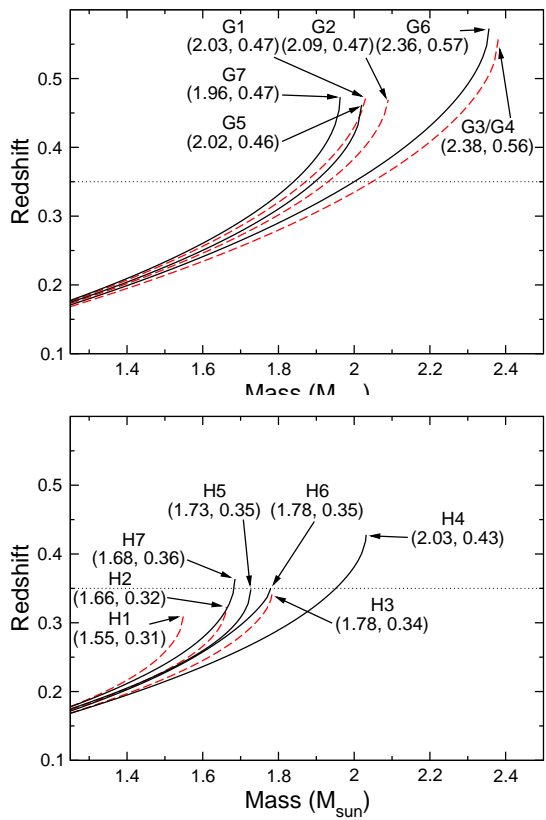

\includegraphics[width=2.8in]{redshiftnoH.eps}
\includegraphics[width=2.8in]{redshiftwithH.eps}
\caption{Gravitational redshift vs.\ mass for relativistic mean field
theory equations of state with hyperons (bottom) and without hyperons (top).
The horizontal line is $z=0.35$ measured for EXO0748-676.
H1 and H2 are ruled out, even allowing for the estimated measurement
errors.
H3, formerly considered the ``stiff'' equation of state, is actually the
softest compatible with the redshift.
H5 through H7 barely satisfy the redshift constraint by construction (see the
next Figure).
}
\label{f:redshiftG}
\end{figure}

\begin{figure}[t]
\includegraphics[width=3.0in]{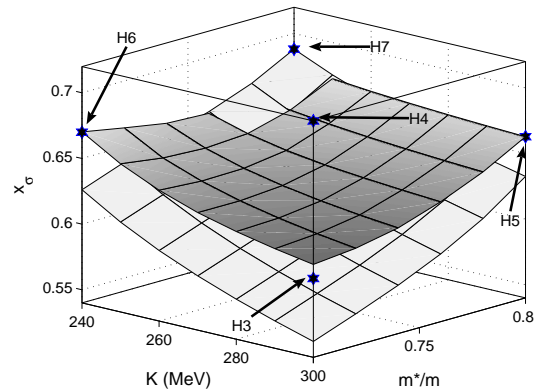}
\caption{
Relativistic mean field theory equations of state can be described by
incompressibility $K$ (in MeV), effective nucleon mass $m^*/m$, and scalar
meson-hyperon coupling $x_\sigma$ (see text).
The dark surface marks those equations of state with a maximum redshift of
0.35.
Equations of state below the surface are incompatible with the observed
redshift of EXO0748-676.
While H3 is below the surface, it is within the estimated 5\% error bar and
should be considered marginally allowed.
Equations of state above the lighter surface are compatible with the 95\%
confidence limit on the mass of Ter~5~I.
The points corresponding to our fiducial equations of state are indicated.
}
\label{f:surface}
\end{figure}

The redshift as a function of mass is plotted for the fiducial relativistic
mean-field equations of state in Fig.~\ref{f:redshiftG} and compared to the
observational constraint from EXO0748-676.
Without hyperons, all equations of state are consistent with $z=0.35$ for
masses greater than the $1.4~M_\odot$ typical of previous measurements,
consistent with the suspected higher masses of accreting stars in x-ray
binaries.
With hyperons, only the stiffest of the usual equations of state (H3) is
marginally consistent with $z=0.35$.
Therefore we favor using H3 as the new ``soft'' equation of state of this
type and H4 as the stiffest.
In fact, varying values of the nuclear incompressibility $K$ and nucleon
effective mass $m^*$ allow for several ``soft'' equations of state marginally
consistent with the redshift, as shown in Fig.~\ref{f:surface}.
Although H3 nominally has $z\le0.34$, it should be considered marginally
consistent because there may be measurement errors of order
5\%~\cite{Bhattacharyya:2004nb} and, more importantly, the maximum redshift
is extremely sensitive to the equation of state at several times nuclear
density.
For example, artificially excluding all hyperons but the $\Lambda$ and
$\Sigma^-$ raises the maximum redshift by 0.03 (a 10\% correction), even though
the populations of those hyperons are very small.

Of the potential-based equations of state, BPAL12 is marginal with
$z\le0.36$ while all the others (including those with hyperons) easily meet
the observational constraint, even when the maximum masses are similar to
the relativistic mean-field theory models.
This is because the maximum-mass stars have 11~km radii in the relativistic
models and 9~km radii in the nonrelativistic ones.
Radii in general are approximately determined by the pressure near nuclear
density~\cite{Lattimer:2000nx}, which in relativistic mean-field theory is
about twice what it is for potential-based models.
Physically this has the simple explanation that most of the matter in the
neutron star is within a factor of two of nuclear density, and so the
pressure at higher densities matters less for the typical radius (though it
is important for the maximum mass).
The fact that the redshift does not constrain the potential-model equations
of state suggests that their low pressure near nuclear density may be
favored (in the sense that there is more unconstrained parameter space).
However, at high densities these models violate causality, which then
favors the relativistic mean field models at high density.

Rather than using a set of fiducial equations of state, one can invert the
problem to ask ``Given an observation of $z=0.35$, what parameters in
relativistic mean-field theory with hyperons are compatible with it?''
As stated in Sec.~II, the three main parameters are $x_\sigma$, $K$, and
$m^*/m$.
If we put them all at the stiff end of their allowed ranges consistent with
nuclear and hypernuclear experiment, we obtain equation of state H4 with a
maximum redshift $z=0.43$.
If we soften the equation of state by lowering $x_\sigma$ (and thus
$x_\omega$ and $x_\rho$) while keeping the other parameters fixed at the H4
values, we get $z=0.35$ at $x_\sigma=0.61$.
Similarly, softening the equation of state by lowering $K$ or raising
$m^*/m$ results in bounds of $K\ge210$~MeV or $m^*/m\le0.84$ respectively,
both of which are less stringent than the experimental
bounds~\cite{Glendenning:1997wn}.
A plot showing the redshift constraint surface (and Ter~5~I mass constraint
surface) in the three-parameter space is shown in Fig~\ref{f:surface}.
The boundaries of the cube correspond to the parameter ranges inferred from
experiment, except for $x_\sigma$ where the experimental lower bound is far
below the redshift constraint surface.
The redshift seems to be most sensitive to the hyperonic coupling parameter
$x_\sigma$, and fairly insensitive to the incompressibility $K$ which is
traditionally considered the measure of stiffness.
This can be seen from the fact that the redshift constraint surface in
Fig.~\ref{f:surface} is fairly flat and that $x_\sigma$ is the only
difference between H3 and H4.
(The mass constraint surface is tilted, \textit{i.e.}\ more correlated with
$K$ and $m^*$.)
The physical explanation is that the main hyperon interaction at high densities
is the repulsion (represented in this framework by the vector meson $\omega$),
and thus increasing the coupling constants decreases the hyperon population of
a given star.
Hyperons, and any other new degrees of freedom, soften the equation of state
at high densities and thus reduce the maximum mass and redshift.

There is also a very recent discussion of a measured $z=0.4$ in the x-ray
binary 4U~1700+24 by Tiengo \textit{et al.}~\cite{Tiengo:2005mm}, but it is
very tentative.
This redshift comes from one spectral line, which is probably better
explained by $z=0.012$ (implying emission well away from the surface of the
star), and there are no other spectral features consistent with $z=0.4$.
H4 would still be compatible with such a redshift, but would be fairly
marginal.

\section{Conclusion}

We have compared equations of state for hyperon stars with new astronomical
observations of mass and gravitational redshift.
Nonrelativistic potential-based models are not greatly constrained by the
new observations.
Relativistic mean-field theory models, however, are tightly constrained by
the observed gravitational redshift.
In fact, the stiffest of these models commonly used in the literature
(which we denote H3) is so soft as to be only marginally compatible with
the observation.
When the full range of parameters consistent with experiments on
hypernuclei is considered, there are still many such equations of state
allowed and the hyperon coupling parameter is found to be the main one
determining the redshift.
As a consequence we advocate that future studies involving these models use
H3 and a new set of canonical parameter values which we denote as equations
of state H4--H7 (see the Appendix).
If through further observations the 95\% confidence limit on the mass of
the neutron star in PSR~J0751+1807 is narrowed to the present 68\%
confidence limit, it would rule out all but H4.

Moving away from the details of a specific model, the general physical
result is this:
The presence of hyperons in neutron stars is constrained but not ruled out
by the gravitational redshift observation (and to a lesser extent by high
mass observations).
In general the equations of state that survive are stiffer than the range
previously considered in the literature.
This means that the hyperons are less numerous, reducing for example the
effect of enhanced cooling and bulk viscosity, which is a subject for
future work~\cite{Nayyar:2005th}.
It may also be useful to consider phenomenological equations of state that
behave like potential-based models near nuclear density but like
relativistic mean-field models at higher densities.

\section{Acknowledgements}

We are grateful to V.~Kalogera and L.~Lindblom for helpful discussions, and
to N.~Glendenning for confirming the typo in
Ref.~\cite{Glendenning:1997wn}.
This work was supported by the National Science Foundation under grants
PHY-0245649 and PHY-0114375 (the Penn State Center for Gravitational Wave
Physics) and by the Penn State Teas program. 

\appendix*
\section{}

For many purposes the detailed microscopic properties of a matter model are
unnecessary, and all that is desired is a tabulation of pressure, energy
density, and baryon number density.
In Tables II, III, IV, V, and VI we provide these for equations of state H3--H7.
The low density ($n<0.3$) parts of some of these (before hyperons or other
strange matter appear) are the same as in Ref.~\cite{Glendenning:1997wn}.
We duplicated the procedure of Glendenning~\cite{Glendenning:1997wn} from
the beginning using constants from Ref.~\cite{Groom:2000in}, which leads to
some discrepancies in the third or fourth significant figure.

\begin{table}[b]
\caption{
Baryon number density $n$ (fm$^{-3}$), energy density $\rho$ (erg/cm$^3$),
and pressure $p$ (dyn/cm$^2$) for H3.}
\begin{ruledtabular}
\begin{tabular}{ccc|ccc}
$n$ & $\rho$ & $p$ &
$n$ & $\rho$ & $p$ \\
\hline
0.03 & 5.041e+13 & 3.581e+31 &
0.63 & 1.221e+15 & 1.842e+35 \\
0.06 & 1.009e+14 & 3.148e+32 &
0.66 & 1.290e+15 & 2.024e+35 \\
0.09 & 1.518e+14 & 1.237e+33 &
0.69 & 1.359e+15 & 2.217e+35 \\
0.12 & 2.031e+14 & 3.124e+33 &
0.72 & 1.430e+15 & 2.407e+35 \\
0.15 & 2.549e+14 & 6.126e+33 &
0.75 & 1.501e+15 & 2.597e+35 \\
0.18 & 3.073e+14 & 1.040e+34 &
0.78 & 1.573e+15 & 2.791e+35 \\
0.21 & 3.604e+14 & 1.624e+34 &
0.81 & 1.646e+15 & 2.991e+35 \\
0.24 & 4.144e+14 & 2.389e+34 &
0.84 & 1.720e+15 & 3.197e+35 \\
0.27 & 4.694e+14 & 3.360e+34 &
0.87 & 1.794e+15 & 3.408e+35 \\
0.30 & 5.263e+14 & 4.261e+34 &
0.90 & 1.870e+15 & 3.596e+35 \\
0.33 & 5.846e+14 & 5.122e+34 &
0.93 & 1.946e+15 & 3.773e+35 \\
0.36 & 6.440e+14 & 6.107e+34 &
0.96 & 2.023e+15 & 3.950e+35 \\
0.39 & 7.044e+14 & 7.236e+34 &
0.99 & 2.100e+15 & 4.125e+35 \\
0.42 & 7.658e+14 & 8.417e+34 &
1.02 & 2.178e+15 & 4.301e+35 \\
0.45 & 8.282e+14 & 9.588e+34 &
1.05 & 2.256e+15 & 4.478e+35 \\
0.48 & 8.914e+14 & 1.083e+35 &
1.08 & 2.336e+15 & 4.657e+35 \\
0.51 & 9.556e+14 & 1.215e+35 &
1.11 & 2.416e+15 & 4.837e+35 \\
0.54 & 1.021e+15 & 1.357e+35 &
1.14 & 2.496e+15 & 5.019e+35 \\
0.57 & 1.087e+15 & 1.508e+35 &
1.17 & 2.577e+15 & 5.202e+35 \\
0.60 & 1.153e+15 & 1.670e+35 &
1.20 & 2.659e+15 & 5.386e+35
\end{tabular}
\end{ruledtabular}
\end{table}

All equations of state but H6 are given up to a baryon density
$n=1.2$~fm$^{-3}$, which is more than sufficient for stable nonrotating
stars.
We stop H6 at $n=0.81$ because at high densities the effective mass of the
proton becomes negative.
This indicates a limitation of the Lagrangian, which was posited as a
low-energy effective theory.
In practice this is not an issue since $n=0.81$ is almost the central
density of the maximum-mass nonrotating H6 star.
We find that extrapolating H6 beyond $n=0.81$ under a wide range of
assumptions only changes the maximum mass of a stable star by 1\%.

\begin{table}[t]
\caption{
Same as the previous Table, but for H4.
}
\begin{ruledtabular}
\begin{tabular}{ccc|ccc}
$n$ & $\rho$ & $p$ &
$n$ & $\rho$ & $p$ \\
\hline
0.03 & 5.041e+13 & 3.581e+31 &
0.63 & 1.242e+15 & 2.416e+35 \\
0.06 & 1.009e+14 & 3.148e+32 &
0.66 & 1.315e+15 & 2.685e+35 \\
0.09 & 1.518e+14 & 1.237e+33 &
0.69 & 1.390e+15 & 2.972e+35 \\
0.12 & 2.031e+14 & 3.124e+33 &
0.72 & 1.466e+15 & 3.260e+35 \\
0.15 & 2.549e+14 & 6.126e+33 &
0.75 & 1.543e+15 & 3.556e+35 \\
0.18 & 3.073e+14 & 1.040e+34 &
0.78 & 1.621e+15 & 3.863e+35 \\
0.21 & 3.604e+14 & 1.624e+34 &
0.81 & 1.701e+15 & 4.182e+35 \\
0.24 & 4.144e+14 & 2.389e+34 &
0.84 & 1.782e+15 & 4.512e+35 \\
0.27 & 4.694e+14 & 3.360e+34 &
0.87 & 1.865e+15 & 4.856e+35 \\
0.30 & 5.256e+14 & 4.561e+34 &
0.90 & 1.949e+15 & 5.191e+35 \\
0.33 & 5.839e+14 & 5.711e+34 &
0.93 & 2.034e+15 & 5.517e+35 \\
0.36 & 6.438e+14 & 6.924e+34 &
0.96 & 2.120e+15 & 5.847e+35 \\
0.39 & 7.050e+14 & 8.313e+34 &
0.99 & 2.208e+15 & 6.183e+35 \\
0.42 & 7.676e+14 & 9.883e+34 &
1.02 & 2.296e+15 & 6.525e+35 \\
0.45 & 8.315e+14 & 1.149e+35 &
1.05 & 2.386e+15 & 6.875e+35 \\
0.48 & 8.967e+14 & 1.321e+35 &
1.08 & 2.476e+15 & 7.232e+35 \\
0.51 & 9.631e+14 & 1.508e+35 &
1.11 & 2.569e+15 & 7.597e+35 \\
0.54 & 1.031e+15 & 1.710e+35 &
1.14 & 2.662e+15 & 7.969e+35 \\
0.57 & 1.100e+15 & 1.928e+35 &
1.17 & 2.756e+15 & 8.346e+35 \\
0.60 & 1.170e+15 & 2.164e+35 &
1.20 & 2.852e+15 & 8.727e+35
\end{tabular}
\end{ruledtabular}
\end{table}

\begin{table}[b]
\caption{
Same as the previous Table, but for H5.
}
\begin{ruledtabular}
\begin{tabular}{ccc|ccc}
$n$ & $\rho$ & $p$ &
$n$ & $\rho$ & $p$ \\
\hline
0.03 & 5.045e+13 & 4.568e+31 &
0.63 & 1.207e+15 & 1.720e+35 \\
0.06 & 1.010e+14 & 2.739e+32 &
0.66 & 1.274e+15 & 1.879e+35 \\
0.09 & 1.518e+14 & 1.123e+33 &
0.69 & 1.343e+15 & 2.045e+35 \\
0.12 & 2.032e+14 & 2.929e+33 &
0.72 & 1.412e+15 & 2.219e+35 \\
0.15 & 2.549e+14 & 5.788e+33 &
0.75 & 1.482e+15 & 2.401e+35 \\
0.18 & 3.073e+14 & 9.754e+33 &
0.78 & 1.552e+15 & 2.591e+35 \\
0.21 & 3.602e+14 & 1.494e+34 &
0.81 & 1.624e+15 & 2.789e+35 \\
0.24 & 4.140e+14 & 2.139e+34 &
0.84 & 1.697e+15 & 2.995e+35 \\
0.27 & 4.686e+14 & 2.916e+34 &
0.87 & 1.770e+15 & 3.209e+35 \\
0.30 & 5.241e+14 & 3.827e+34 &
0.90 & 1.844e+15 & 3.431e+35 \\
0.33 & 5.806e+14 & 4.874e+34 &
0.93 & 1.919e+15 & 3.660e+35 \\
0.36 & 6.388e+14 & 5.840e+34 &
0.96 & 1.995e+15 & 3.898e+35 \\
0.39 & 6.982e+14 & 6.819e+34 &
0.99 & 2.071e+15 & 4.136e+35 \\
0.42 & 7.586e+14 & 7.883e+34 &
1.02 & 2.149e+15 & 4.377e+35 \\
0.45 & 8.199e+14 & 9.040e+34 &
1.05 & 2.227e+15 & 4.623e+35 \\
0.48 & 8.822e+14 & 1.029e+35 &
1.08 & 2.306e+15 & 4.875e+35 \\
0.51 & 9.454e+14 & 1.156e+35 &
1.11 & 2.386e+15 & 5.132e+35 \\
0.54 & 1.010e+15 & 1.287e+35 &
1.14 & 2.466e+15 & 5.396e+35 \\
0.57 & 1.074e+15 & 1.424e+35 &
1.17 & 2.548e+15 & 5.666e+35 \\
0.60 & 1.140e+15 & 1.558e+35 &
1.20 & 2.630e+15 & 5.942e+35
\end{tabular}
\end{ruledtabular}
\end{table}

\begin{table}[t]
\caption{
Same as the previous Table, but for H6.
}
\begin{ruledtabular}
\begin{tabular}{ccc|ccc}
$n$ & $\rho$ & $p$ &
$n$ & $\rho$ & $p$ \\
\hline
0.03 & 5.038e+13 & 2.458e+31 &
0.45 & 8.220e+14 & 8.987e+34 \\
0.06 & 1.009e+14 & 3.597e+32 &
0.48 & 8.846e+14 & 1.026e+35 \\
0.09 & 1.517e+14 & 1.380e+33 &
0.51 & 9.481e+14 & 1.166e+35 \\
0.12 & 2.031e+14 & 3.308e+33 &
0.54 & 1.013e+15 & 1.320e+35 \\
0.15 & 2.549e+14 & 6.183e+33 &
0.57 & 1.078e+15 & 1.489e+35 \\
0.18 & 3.073e+14 & 1.001e+34 &
0.60 & 1.145e+15 & 1.674e+35 \\
0.21 & 3.603e+14 & 1.513e+34 &
0.63 & 1.212e+15 & 1.875e+35 \\
0.24 & 4.139e+14 & 2.160e+34 &
0.66 & 1.281e+15 & 2.090e+35 \\
0.27 & 4.685e+14 & 2.974e+34 &
0.69 & 1.350e+15 & 2.299e+35 \\
0.30 & 5.241e+14 & 3.933e+34 &
0.72 & 1.421e+15 & 2.512e+35 \\
0.33 & 5.816e+14 & 4.709e+34 &
0.75 & 1.493e+15 & 2.732e+35 \\
0.36 & 6.402e+14 & 5.588e+34 &
0.78 & 1.565e+15 & 2.959e+35 \\
0.39 & 6.998e+14 & 6.621e+34 &
0.81 & 1.639e+15 & 3.188e+35 \\
0.42 & 7.604e+14 & 7.807e+34 &
& & 
\end{tabular}
\end{ruledtabular}
\end{table}

\begin{table}[b]
\caption{
Same as the previous Table, but for H7.
}
\begin{ruledtabular}
\begin{tabular}{ccc|ccc}
$n$ & $\rho$ & $p$ &
$n$ & $\rho$ & $p$ \\
\hline
0.03 & 5.039e+13 & 3.353e+31 &
0.63 & 1.195e+15 & 1.621e+35 \\
0.06 & 1.009e+14 & 3.585e+32 &
0.66 & 1.261e+15 & 1.777e+35 \\
0.09 & 1.518e+14 & 1.329e+33 &
0.69 & 1.328e+15 & 1.942e+35 \\
0.12 & 2.031e+14 & 3.151e+33 &
0.72 & 1.396e+15 & 2.116e+35 \\
0.15 & 2.549e+14 & 5.842e+33 &
0.75 & 1.465e+15 & 2.298e+35 \\
0.18 & 3.072e+14 & 9.426e+33 &
0.78 & 1.535e+15 & 2.488e+35 \\
0.21 & 3.601e+14 & 1.401e+34 &
0.81 & 1.605e+15 & 2.715e+35 \\
0.24 & 4.136e+14 & 1.966e+34 &
0.84 & 1.677e+15 & 2.897e+35 \\
0.27 & 4.678e+14 & 2.645e+34 &
0.87 & 1.749e+15 & 3.114e+35 \\
0.30 & 5.228e+14 & 3.442e+34 &
0.90 & 1.822e+15 & 3.339e+35 \\
0.33 & 5.787e+14 & 4.362e+34 &
0.93 & 1.896e+15 & 3.572e+35 \\
0.36 & 6.356e+14 & 5.349e+34 &
0.96 & 1.971e+15 & 3.812e+35 \\
0.39 & 6.942e+14 & 6.252e+34 &
0.99 & 2.047e+15 & 4.059e+35 \\
0.42 & 7.537e+14 & 7.227e+34 &
1.02 & 2.124e+15 & 4.309e+35 \\
0.45 & 8.140e+14 & 8.292e+34 &
1.05 & 2.201e+15 & 4.564e+35 \\
0.48 & 8.752e+14 & 9.454e+34 &
1.08 & 2.279e+15 & 4.825e+35 \\
0.51 & 9.374e+14 & 1.071e+35 &
1.11 & 2.358e+15 & 5.093e+35 \\
0.54 & 1.000e+15 & 1.199e+35 &
1.14 & 2.438e+15 & 5.369e+35 \\
0.57 & 1.064e+15 & 1.332e+35 &
1.17 & 2.519e+15 & 5.651e+35 \\
0.60 & 1.129e+15 & 1.473e+35 &
1.20 & 2.601e+15 & 5.941e+35
\end{tabular}
\end{ruledtabular}
\end{table}

\bibliography{paper}

\end{document}